\documentclass[
    ,final            
  ]
  {aipproc}

\layoutstyle{6x9}

\begin{document}

\title{How is the Diamagnetic Effect (DME) Relevant to Stellar Surface Phenomena? }


\author{Netzach Farbiash}{
  address={Ben-Gurion University, Beer-Sheva 84105, Israel}
}

\author{Raphael Steinitz}{
  address={Ben-Gurion University, Beer-Sheva 84105, Israel}
}

\begin{abstract}
 The structure of stellar atmospheres can be modified if they include embedded diverging magnetic fields.
 This is due to the presence of speed filters. Here we examine some of the relevant effects of speed filters,
 using numerical simulations of the motion of charged particles.
 We introduce the concept of "floating" particles and point out the relevance of the
 diamagnetic effect to the evolution of Maxwellian velocity distributions into non-Maxwellian distributions.

\end{abstract}

\maketitle


\section{Introduction}

The motion of charged particles in the presence of magnetic fields
is a fundamental problem in astrophysics. Magnetic fields play
significant roles in stellar atmospheres as well as in outer space.
Diverging (geometrically) magnetic fields are common in the
universe, so that it is essential to understand the consequences of
the motion of a charged particle in such fields when also embedded
in a gravitational field.

For the sake of simplicity as well as to obtain analytic solution,
we mimic here the true magnetic field structures by using monopoles.
For sufficiently small regions, a monopole field is a reasonable
approximation to the true field (for example, field lines of a
magnetic dipole, can be locally presented as filed lines of a
magnetic monopole (see fig 1)).

\begin{figure}[!ht]
  \includegraphics[height=.3\textheight]{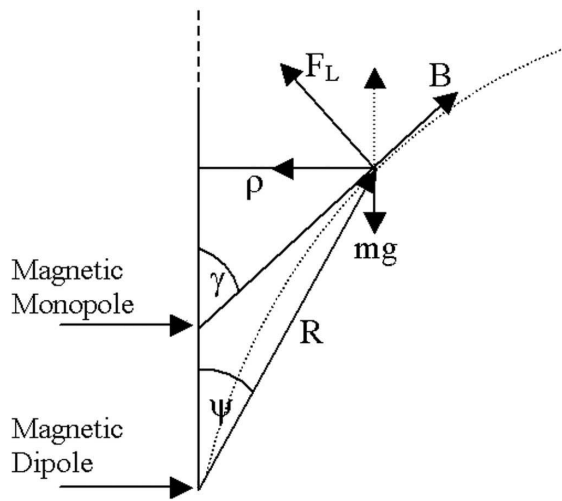}
  \caption{Dipole field approximated locally by a magnetic monopole.
}
\end{figure}

\section{Speed Filters and Floating Particles}

We follow the motion of a charged particle, located at a height
$z_{0}$ above a magnetic monopole in the presence of a gravitational
field. The height $z_{0}$ is chosen so that the vertical component
of the Lorentz force is equal and opposite to the gravitational
force $\left(\sum F_{z}=0\right)$. This choice ensures that the
particle is "floating" (see fig. 2). The initial conditions:

\begin{equation}
|F_{Lorentz}|\sin\alpha=mg
\end{equation}

\begin{equation}
|F_{Lorentz}|\cos\alpha=\frac{mv^{2}_{\perp}}{r}
\end{equation}

\noindent From the geometry of the problem, we obtain that
$$\tan\alpha=\frac{r}{z_{0}},$$
 and from eqs. 1 and 2, we get
$$\tan\alpha=\frac{v^{2}_{\perp}}{gr} .$$

\noindent Thus,

\begin{equation}
z_{0}=\frac{v^{2}_{\perp}}{g}  .
\end{equation}

\begin{figure}[!ht]
  \includegraphics[height=.3\textheight]{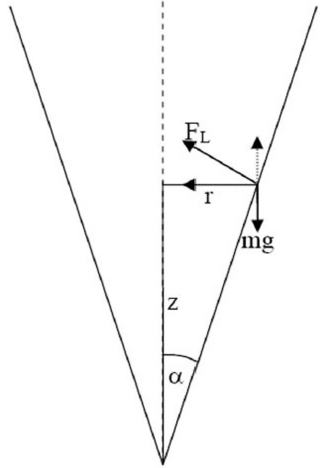}
  \caption{Speed filter.
}
\end{figure}

We see from eq. (3) that at a height $z_{0}$ above the monopole, the
"floating height" depends on $v^{2}_{\perp}$. If the particle has a
velocity component parallel to component to the field, it will
oscillate around $z_{0}$. For $v^{2}_{\perp}<gz_{0}$ the  particle
falls and the diamagnetic effect (DME) increases until the vertical
component of the Lorentz force is equal to the gravitational force.
In this case, we have $z_{down}<z_{0}$. For $v^{2}_{\perp}>gz_{0}$
the particle rises and the DME decreases until the the vertical
component of the Lorentz force is equal to the gravitational force.
In that case, we have $z_{up}>z_{0}$. We conclude that the
combination of the DME and gravitation can operate as a speed
filter.

\section{Numerical simulation}

The magnetic monopole field is

\begin{equation}
\overrightarrow{B}=\frac{\beta}{R^{3}}\overrightarrow{R}     ,
\end{equation}

\noindent where $\beta$ is a constant.

\noindent The equation of motion of a particle (mass m, charge q) in
the presence of gravitation is

\begin{equation}
\ddot{\vec{R}}=\left(\frac{q\beta}{m}\right)\frac{1}{R^{3}}\left(\dot{\vec{R}}\times\vec{R}\right)+\vec{g}.
\end{equation}

We follow the position of a charged particle (proton) as function of
time, in the presence of gravitation alone as well as in the
presence of both gravitation and the DME. The particle's initial
position is chosen to be at a given height above the magnetic
monopole.

\begin{figure}[!ht]
  \includegraphics[height=.3\textheight]{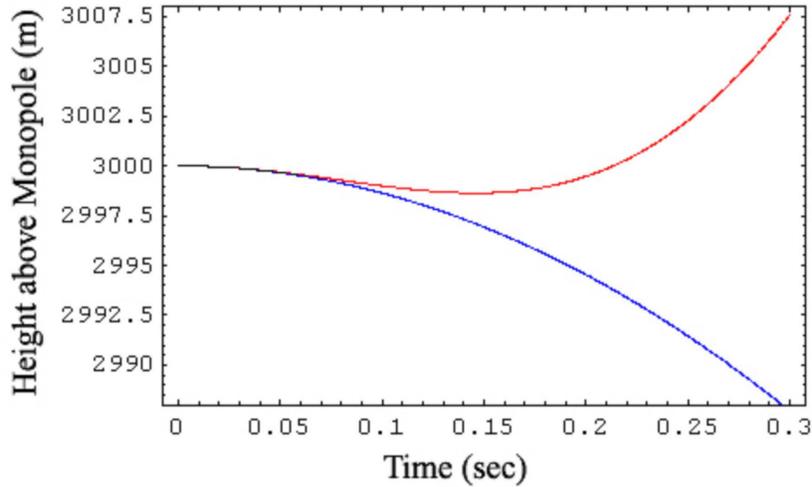}
  \caption{Height of proton as a function of time with both diamagnetic effect and gravity (red line),
  and with gravity only (blue line).
Initial conditions are: $V_{0x}=0.5\sqrt{3Kb\cdot T/m_{p}}$ ,
$V_{0y} = 0.5\sqrt{3Kb\cdot T/m_{p}}$ ,  $V_{0z} = 0$ ,  $Z_{0} =
3\cdot10^{3} m$ ,  and $T=5\cdot10^{3} K$}
\end{figure}

\begin{figure}[!ht]
  \includegraphics[height=.3\textheight]{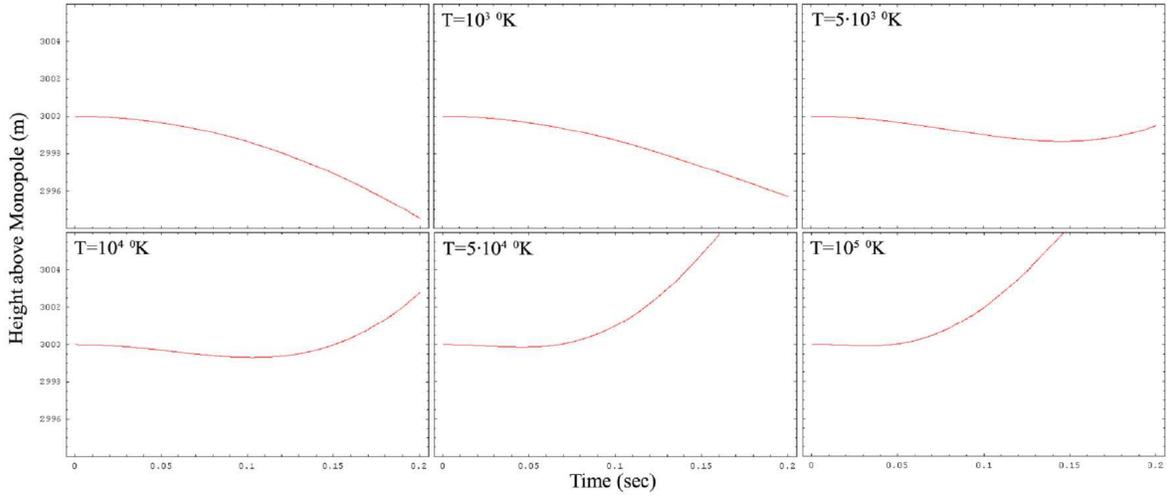}
  \caption{Height of proton as a function of time. Initial conditions,
  are the same for all graphs, except that of the temperature: $V_{0x}=0.5\sqrt{3Kb\cdot T/m_{p}}$ ,
$V_{0y} = 0.5\sqrt{3Kb\cdot T/m_{p}}$ ,  $V_{0z} = 0$ ,  $Z_{0} =
3\cdot10^{3} m$ ,  and $T=5\cdot10^{3} K$. The first frame (top
left) shows height of protons as function of time in the presence of
gravity alone.}
\end{figure}

\begin{figure}[!ht]
  \includegraphics[height=.3\textheight]{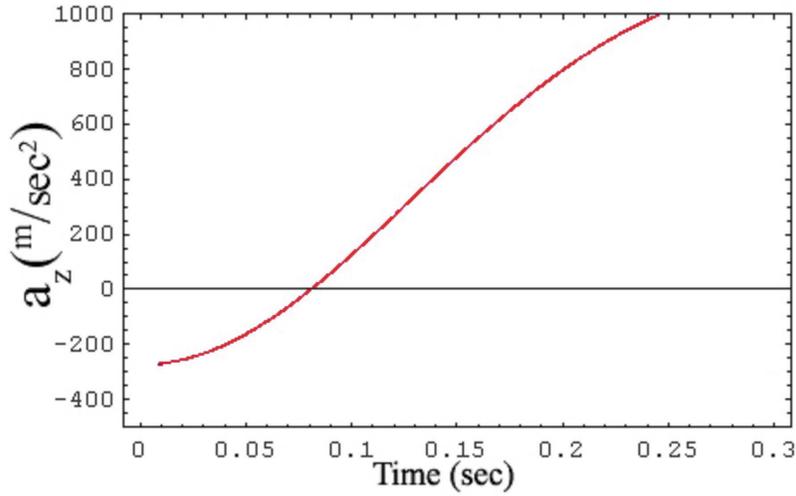}
  \caption{Acceleration along z axis of a proton as a function of time with diamagnetic effect and
  gravity.
  Initial conditions are: $V_{0x}=0.5\sqrt{3Kb\cdot T/m_{p}}$ ,
$V_{0y} = 0.5\sqrt{3Kb\cdot T/m_{p}}$ ,  $V_{0z} = 0$ ,  $Z_{0} =
3\cdot10^{3} m$ ,  and $T=5\cdot10^{3} K$}
\end{figure}

\section{Results}

The above results are depicted in Figs. 3 and 4, while in Fig. 5,
the acceleration is also shown. In the presence of the DME as well
as gravitation, particles fall more slowly than in a gravitation
field alone or can even rise (Fig. 3). We see from Fig. 5, that the
effect of the DME increases in importance as $v^{2}_{\bot}$ (i.e.
temperature) grows, as expected.

\section{Conclusions}

The simulations demonstrate that the hydrostatic structure of an
atmosphere is modified by the DME. The values of pressure and
temperature change, essentially because the "effective" gravity is
diminished (or inverted) selectively by the faster particles of the
normal distribution. One can expect significant changes in the
ionization equilibrium. The restructuring of the atmosphere can
cause "apparent peculiarities" in the spectra of stars. The
relevance of the DME to stellar atmospheric structure is, thus, due
to the modification of effective gravity, which is particle speed
dependence. The essence of the process is the speed filter.

In the future we will include collisions in order to follow the
evolution of a Maxwellian distribution into a non-Maxwellian one, as
required for the initial conditions in the theory discussed by
Scudder (1992a, 1992b, 1994), and Anderson (1994).

\vspace{2\baselineskip} \noindent We thank S. Owocki for useful
discussions and helpful suggestions.

\bibliographystyle{aipprocl} 

\end{document}